\documentclass[]{interact}

\usepackage{epstopdf}
\usepackage[caption=false]{subfig}

\usepackage[numbers,sort&compress]{natbib}
\bibpunct[, ]{[}{]}{,}{n}{,}{,}
\renewcommand\bibfont{\fontsize{10}{12}\selectfont}
\makeatletter
\def\NAT@def@citea{\def\@citea{\NAT@separator}}
\makeatother

\theoremstyle{plain}

\theoremstyle{definition}

\theoremstyle{remark}

\usepackage[hidelinks=true]{hyperref}

\usepackage{pgf, tikz}
\usetikzlibrary{arrows, automata}
\usetikzlibrary{shapes.geometric}
\usepackage{pgfplots}
\pgfplotsset{compat=1.17}
\usetikzlibrary{shapes,arrows,shadows,decorations.pathreplacing}

\tikzstyle{square}=[draw]        

\tikzstyle{mybox}=[draw=gray!95, fill=white, very thick,
                     rectangle, inner sep=7pt, inner ysep=7pt]

\tikzstyle{mybox2}=[draw=gray!50, fill=gray!20, very thick, 
                     rectangle, inner sep=7pt, inner ysep=7pt]

\usepackage{pifont}
\newcommand{\cmark}{\text{\ding{51}}}%
\newcommand{\xmark}{\text{\ding{55}}}%

\usepackage{amsmath}
\usepackage{amsfonts}
\usepackage{amssymb}
\usepackage{pdfpages}
\usepackage{bm} 

\begin{document}

\articletype{ARTICLE}

\title{Multivariate generalized linear mixed models for underdispersed count data}

\author{
\name{Guilherme Parreira da Silva\textsuperscript{a}\thanks{CONTACT Guilherme Parreira da Silva. Email: guilhermeparreira.silva@gmail.com}, Henrique Aparecido Laureano\textsuperscript{a}, Ricardo Rasmussen Petterle\textsuperscript{b}, Paulo Justiniano Ribeiro J\'{u}nior\textsuperscript{a}, Wagner Hugo Bonat\textsuperscript{a}}
\affil{\textsuperscript{a}Laboratory of Statistics and Geoinformation, Department of Statistics, Paran\'{a} Federal University, Curitiba, Brazil; \textsuperscript{b}Department of Integrative Medicine, Paran\'{a} Federal University, Curitiba, Brazil}
}

\maketitle

\begin{abstract}
Researchers are often interested in understanding the relationship between a set of covariates and a set of response variables. To achieve this goal, the use of regression analysis, either linear or generalized linear models, is largely applied. However, such models only allow users to model one response variable at a time. Moreover, it is not possible to directly calculate from the regression model a correlation measure between the response variables. In this article, we employed the Multivariate Generalized Linear Mixed Models framework, which allows the specification of a set of response variables and calculates the correlation between them through a random effect structure that follows a multivariate normal distribution. We used the maximum likelihood estimation framework to estimate all model parameters using Laplace approximation to integrate out the random effects. The derivatives are provided by automatic differentiation. The outer maximization was made using a general-purpose algorithm such as \texttt{PORT} and \texttt{BFGS}. We delimited this problem by studying only count response variables with the following distributions: Poisson, negative binomial (NB) and COM-Poisson. While the first distribution can model only equidispersed data, the second models equi and overdispersed, and the third models all types of dispersion. The models were implemented on software \texttt{R} with package \texttt{TMB}, based on \texttt{C++} templates. Besides the full specification, models with simpler structures in the covariance matrix were considered (fixed and common variance, fixed dispersion, $\rho$ set to 0). These models were applied to a dataset from the National Health and Nutrition Examination Survey, where three underdispersed response variables were measured at 1281 subjects. The COM-Poisson model full specified overcome the other two competitors considering three goodness-of-fit indexes: AIC, BIC and likelihood. As a consequence, it estimated parameters with smaller standard error and a greater number of significant correlation coefficients. Therefore, the proposed model can deal with multivariate count responses and measures the correlation between them taking into account the effects of the covariates.
\end{abstract}


\section{Introduction}

Researchers are often interested to understand the relationship between a set of response variables and a set of covariates in their different areas of study. For example, doctors are interested to know whether polypharmacy is related to complications after surgery; veterinarians may be interested to know whether animal welfare is related to meat quality; administrators may be interested to know whether the usage of the new policy has improved social indicators. When we have a set of covariates and one specific objective (response variable), these situations can be addressed in the statistical literature by regression models. Certainly, one of the most known and widely applied is the linear regression (LM) model \cite{linearmodelsorigin}. 

LM is widely used due to its simplicity and the general ordinary least squares estimation procedure, which is covered in different textbooks in different areas such as business, numerical optimization, agronomy, among others. To correct apply it, it is necessary to verify whether the residuals are independent, not autocorrelated and with homogeneous variance. These assumptions can be too restrictive depending on the context. 

As an alternative, the generalized linear model (GLM) is a more flexible approach \cite{glm}. GLM is a class of models that generalizes LM by supporting response variables that belong to the exponential family. The distributions that belong to the exponential family are binomial, gamma, inverse Gaussian, normal, and Poisson. Moreover, it is built upon a link function that connects the linear predictor to the expectation of the response variable, and its variance can be related to its mean.

A count data represents the number of times that an event occurs in a fixed interval, such as time, space, distance, area, among others. Therefore, it is finite and non-negative. One example of such data is the number of times ear, body posture and head orientation changes in ewes after brushing (a treatment proposed to increase welfare in animals) \cite{priscillamanyresp}. The Poisson distribution is widely used for this purpose but relies on the fact that the variance of the data is equal to its mean, which is known as equidispersion. However, this assumption is too restrictive, and a different mean-variance relationship can be found, such as overdispersion and underdispersion. Overdispersion occurs when the variance of the data is greater than the mean, and it is often found in practice. It usually happens due to excess of zeros, heavy-tailed distribution, or absence of a covariate to model the data \cite{overdispersion}. On the other hand, underdispersion occurs when the variance is smaller than the mean and it is less usual than the overdispersion case \cite{gammacount}. An underdispersed random variable is characterized by a smaller range of observed counts compared to an overdispersed count data.

Different distributions have been proposed to model count data. When the variance is equal to the mean, after considering the effects of all covariates, Poisson is the most obvious choice. When the data are overdispersed, negative binomial (NB) type II under the same framework of GLM is a good choice. The Extended Poisson Tweedie \cite{extendedPT} based on the Poisson Tweedie distribution \cite{pt1,pt2}, Conway-Maxwell-Poisson (COM-Poisson) \cite{compoisson} and Gamma Count \cite{gammacount} distributions can be used to model either under-, equi- or overdispersed data. The drawback of them is that the probability mass function (pmf) does not have a closed-form expression, making the process of inference time consuming for procedures that rely on the pmf, such as likelihood.

Beyond different distributions, hurdle and the zero-inflated models \cite{hurdlezi} can be used to model count data, especially when the data is zero-inflated \cite{zeroinflated}. Two widely-known alternatives are the hurdle and the zero-inflated models. The disadvantage of them is that interpretability becomes cumbersome especially for hurdle model. 

All the strategies pointed out here consider that we have available only one response variable. However, it is not difficult to find in the literature datasets where the researchers possess more than 1 response variable for the same study. Usually, the analysis is made by each response individually due to the lack of alternatives in statistical software. Nevertheless, there is an increasing interest in the literature to develop models or distributions that can handle multivariate responses, that is, when there is more than one response variable \cite{wbonat.article}.

One approach to model more than one response variable simultaneously is to construct multivariate distributions for count data. \cite{multipoisson} present three alternatives to model multivariate count data. The first assumes that the marginal distribution is Poisson, and a multivariate distribution is build under copulas or multivariate distribution theory \cite{multipoissontheory}. The second uses a mixture of independent Poisson. The third method generalizes the first one, where the conditional distributions are also Poisson. However, none of them deals with under-dispersed data. \cite{multgenpoisregmodel} proposed a multivariate generalized Poisson regression model based on the multivariate generalized Poisson distribution (MGPD) that can deal with equi-, under- or overdispersed data, the correlation estimates can either be positive or negative, and the estimation is made under the maximum likelihood (ML) framework. 


\cite{winkelmannbook} provides an overview of different distributions/models for count data. It presents the multivariate NB model (MNBM) and the multivariate Poisson-gamma mixture model (MPGM), which allow only for overdispersion and nonnegative correlation. It also presents the multivariate Poisson-lognormal regression (MPLR) model and the latent Poisson-normal regression model, which allows positive and negative correlation. However, it is suitable only for overdispersed data.


\cite{wbonat.pkg} proposed the Multivariate Covariance Generalized Linear Models (MCGLM), which is a class of models based on quasi-likelihood \cite{quasilikelihood} and generalized estimating equations (GEE) \cite{gee}, that allows fitting multivariate models using only second-order moments assumptions with correlated data. Bayesian Regression Models using Stan - \text{brms} package \cite{brms} and MCMC Generalized Linear Mixed Models - \text{MCMCglmm} package \cite{mcmcglmm} provide a framework to model multivariate models via Bayesian inference \cite{bayesianinference}.

Another alternative is to model the correlation between response variables for the same individual using the class of hierarchical GLM \cite{hglm}. This class allows to model correlated variables or individuals via a random effect, an unobserved variable, that can follow any distribution. When the distribution of the random effect is Gaussian, we have the Generalized Linear Mixed Models (GLMM). However, GLMM is widely known and used to model correlation between sample units, not for response variables, such methodology is implemented in consolidated packages in software \texttt{R} \cite{rsoft}, such as, glmmTMB \cite{glmmTMB}, lme4 \cite{lme4} and nlme \cite{nlme}.

In this article, we propose to model multivariate underdispersed count data under the framework of GLMMs to accommodate correlation between response variables. 

This article contains six sections including this introduction. Section 2 describes the dataset used as a model example of application. Section 3 presents a literature review of the GLMM model. Section 4 proposes the MGLMM model along with the estimation procedure. Section 5 presents the results of the model applied to the data from Section 2. Finally, Section 6 discusses the main contributions of this article and future work is also pointed.

  
\section{DATASET: NATIONAL HEALTH AND NUTRITION EXAMINATION SURVEY}
\label{cap:nhanes}

  
The National Health and Nutrition Examination Survey (NHANES) is a program that studies the health and nutrition status of adults and children in the United States. This survey is being conducted every year since the early 1960s. Nowadays, it examines a nationally representative sample of about 5,000 subjects each year \cite{nhanes}. Among different types of data collected, the main objective of this analysis was to investigate whether demographical variables influence sexual behaviour.

We used the sex behaviour and demographic datasets. From the first, it was selected three response variables Nmsp (Number of male sex partners in the past year), Nmosp (Number of male oral sex partners in the past year) and Nspfy (Number of sex partners who are five years older in the past year). From the second dataset, it was used the covariates race (0 = Others, 1 = White), education level (range from 1 = Less Than $9^{\text{th}}$ Grade to 5 = College Graduate or above) and marital status (0 = Marital, 1 = Others). After deleting those respondents who had missing data, the sample consisted of 1281 women, with ages ranging from 18 to 80, 57\% were married, 43\% white and 31\% had some college or associates (AA) degree.

One way to analyse this dataset is via regression models. Once all three response variables are count data, we need to choose a suitable probability mass function for this data. In this paper, we compared the Poisson, NB and COM-Poisson probability distributions. As this is a cross-sectional study, we do not have responses correlated over time. However, we may have correlated responses variables, once they were measured in the same individual. The model proposed in \autoref{cap:modelo} can accommodate this situation. 

\autoref{tab:desc_nhanes} presents the mean, variance, Fisher Dispersion Index (DI) \cite{dispersionindex} for every response variable and the generalized dispersion index (GDI) \cite{generalizedDI} for the dataset. The main reason for choosing those response variables is the fact that Nmsp and Nmosp can be considered marginally underdispersed, once the sample variance is smaller than the sample mean. Moreover, Nspfy is only a little overdispersed. It is easily seen from the DI, which is calculated by dividing the variance of the variable by its mean. A variable with $\text{DI}>1$ can be said as overdispersed, DI = 1 as equidispersed and $\text{DI}<1$ underdispersed.  

Moreover, the GDI is a recently proposed multivariate dispersion index. It is calculated based on the expectation of each variable and the covariance between them. When the number of variables is equal to 1, it is just the classical Fisher DI. A $\text{GDI}>1$ classifies the multivariate responses as overdispersed, GDI = 1 as equidispersed and $\text{GDI}<1$ underdispersed. The standard error (SE) for GDI is calculated using the asymptotic behaviour of the estimator. According to this index, a 95\% confidence interval based on SE suggests that this dataset is equidispersed as 1 is included on it.

\begin{table}
\caption{DESCRIPTIVE MEASUREMENTS FOR NHANES RESPONSE VARIABLES}
\centering
\begin{tabular}[t]{lccccccc}
\toprule
\multicolumn{1}{c}{ } & \multicolumn{3}{c}{Spearman Correlation $\rho$} & \multicolumn{1}{c}{Mean} & \multicolumn{1}{c}{Variance} & \multicolumn{1}{c}{DI} & \multicolumn{1}{c}{GDI(SE)} \\
\cmidrule(l{3pt}r{3pt}){2-4} \cmidrule(l{3pt}r{3pt}){5-5} \cmidrule(l{3pt}r{3pt}){6-6} \cmidrule(l{3pt}r{3pt}){7-7} 
& Nmsp & Nmosp & Nspfy &  &  &  & \\
\cmidrule(l{3pt}r{3pt}){1-7}
Nmsp &  & 0.033 & 0.222 & 1.313 & 1.084 & 0.826 & \\
Nmosp &  &  & 0.038 & 0.372 & 0.350 & 0.939 & 1.092(0.22)\\
Nspfy &  &  &  & 0.368 & 0.411 & 1.117 & \\
\bottomrule
\end{tabular}
\label{tab:desc_nhanes}
\end{table}

\begin{figure}
\vspace{0.35cm}
\setlength{\abovecaptionskip}{.0001pt}
\centering
\includegraphics[width=0.95\textwidth]{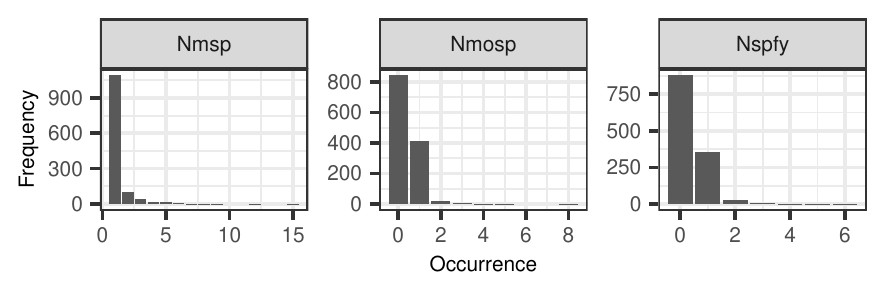}
\caption{BARPLOT FOR EACH RESPONSE VARIABLE FROM NHANES DATA}\label{fig:descbarnhanes}
\end{figure}

The three response variables show no or small (Nmsp and Nspfy) correlation between them. \autoref{fig:descbarnhanes} shows the barplot of each response variable. We can see that there is a higher frequency for non-occurrence of events, rather than the occurrence.

  %
\section{GENERALIZED LINEAR MIXED MODELS (GLMM)}
\label{cap:glmm}
%
  
The GLMM class can be used to address dependent observations, overdispersion, among others \cite{glmm}. In addition to the fixed effect from the GLM, it accommodates extra variability through an unobservable variable, the random effect, and is calculated for each set of dependent observations. Once the random effect is an unobserved random variable, we can write the GLMM in a hierarchical structure. The following notation is valid when the dependent observations come for repeated measurements and longitudinal studies:
\begin{align}
\label{eq:glmm}
\mathrm{Y}_{il}\mid \mathrm{\mathbf{b}}_{i\cdot} \sim f(\mu_{il}, \phi) \nonumber \\
\mathrm{g}(\mu_{il}) = x_{ij}^{\top}\boldsymbol{\beta} + z_{ij}^{\top}\mathrm{\mathbf{b}}_{i\cdot} \\
\mathrm{\mathbf{b}}_{i\cdot} \sim \mathcal{N}\left ( \mathbf{0},\boldsymbol{\Sigma} \right ), \nonumber
\end{align}
where $\mathrm{Y}_{il}$ is the i-th unit sample measured in the l-th times (in a longitudinal study) or in the l-th group (in a repeated measurements study), $\mathrm{\mathbf{b}}_{i\cdot}$ is a vector of random effects for each sample unit that follows a multivariate normal distribution with mean $\mathbf{0}$ and covariance matrix $\boldsymbol{\Sigma}$ to accommodate the variance of each random effect and covariance among them, $f$ is the conditional distribution of $\mathrm{Y}_{il}$ on $\mathrm{\mathbf{b}}_{i\cdot}$, $\mu_{il}$ is the mean for every sample unit $i$ and measurement $l$, $\mathrm{g}(.)$ is a suitable link function, and $\mathrm{z}_{ij}^{\top}$ is the design matrix associated to the random effect.


\section{MULTIVARIATE GENERALIZED LINEAR MIXED MODEL (MGLMM) FOR COUNT DATA}
\label{cap:modelo}


Let $\mathrm{Y}_{ir}$ be the multivariate outcome for subject \emph{i}, where $i = 1,\dots,n$ and response variable \emph{r}, where $r = 1,\dots,k$. Suppose that a set of \emph{p} known covariates is available for each response \emph{r}, therefore, $\mathrm{x}_{irj}$ is the value of the \emph{j-th} covariate for individual \emph{i} and response \emph{r}. The purpose of this article is to provide a joint model for a set of response variables. We start from the standard GLMM model specification with only a random intercept. The first component we need to assume is the conditional distribution of the response variables:

\begin{equation*}
\mathrm{Y}_{ir} \mid \mathrm{b}_{ir} \sim f(\mu_{ir};\phi_{r}),
\end{equation*}
where all response variables share the same distribution. However, with different mean and dispersion parameters. The second component we need to specify is the linear predictor:

\begin{equation*}
\label{eq:linear_predictor}
\mathrm{g}_r(\mu_{ir}) = x_{irj}^{\top}\boldsymbol{\beta}_{r} + \mathrm{b}_{ir},
\end{equation*}
where $\mathrm{g}_r(.)$ is a suitable link function, $\boldsymbol{\beta}_{r}$ is a $p \times 1$ vector of parameter estimates and $\mathrm{b}_{ir}$ is the random intercept value for each individual and response variable. Lastly, the distribution of the random effects is specified by:

\begin{equation*}
\begin{pmatrix}
\mathrm{b}_{i1}\\ 
\mathrm{b}_{i2}\\ 
\vdots \\ 
\mathrm{b}_{ir}
\end{pmatrix}
\sim \mathrm{NM} \begin{pmatrix}
\begin{bmatrix}
0\\ 
0\\ 
\vdots\\ 
0
\end{bmatrix};
\underset{r\times r}{\boldsymbol{\Sigma}} = \begin{bmatrix}
\sigma^{2}_{1} & \rho_{12}\sigma_1\sigma_2 & \dots & \rho_{1r}\sigma_1\sigma_r\\ 
\rho_{21}\sigma_2\sigma_1 & \sigma^{2}_{2}  & \dots & \rho_{2r}\sigma_2\sigma_r \\ 
\vdots & \vdots & \ddots  & \vdots \\ 
\rho_{r1}\sigma_r\sigma_1 & \rho_{r2}\sigma_r\sigma_2 & \dots & \sigma^{2}_{r}
\end{bmatrix}
\end{pmatrix}
,
\end{equation*}
where each random effect has mean 0, variance $\sigma^2$, and $\rho_{r{r}'} (r\neq{r}')$ measures the correlation between each pair of random effects. Despite this is a general modelling framework, this article addresses only the same distribution (either Poisson, binomial negative or COM-Poisson) and link function (logarithm) for all responses variables. 

An important question for this type of model is whether it is possible to estimate simultaneously the dispersion parameters of the pmf and the variance parameters of the random effects, once they measure variability related to the same random variable, and can cause identifiability problems.
%
\subsection{INFERENCE AND ESTIMATION}
\label{cap:estimacao}
%
%
%
In this section, we present the estimation procedure used to obtain the parameter estimates based on the likelihood function. In order to estimate the model presented in \autoref{cap:modelo} under the maximum likelihood (ML) paradigm, it is necessary to construct the joint distribution for both random variables ($\mathrm{Y}_{ir} \text{ and } \mathrm{b}_{ir}$). Once we have a hierarchical structure, the joint distribution of $\mathrm{Y}_{ir} \text{ and } \mathrm{b}_{ir}$ can be factored into $f(\mathrm{Y}_{ir},\mathrm{b}_{ir}) = f(\mathrm{Y}_{ir}\mid\mathrm{b}_{ir})f(\mathrm{b}_{ir})$. As only $\mathrm{Y}$ is observed, we are interested on the marginal distribution of $\mathrm{Y}$ that can be obtained integrating out $\mathrm{b}_{ir}$, that is, $f(\mathrm{Y}_{ir}) = \int f(\mathrm{Y}_{ir}\mid\mathrm{b}_{ir})f(\mathrm{b}_{ir})\mathrm{db}_{ir}$, which in practical terms means that the joint distribution is averaged over the $\mathrm{b}_{ir}$ terms.

Now, we particularize the ML estimation method to the model described in \autoref{cap:modelo}. The objective is to estimate the parameter vector $\boldsymbol{\theta} = \{\boldsymbol{\beta}, \boldsymbol{\phi}, \boldsymbol{\sigma^2}, \boldsymbol{\rho}\}$, where $\boldsymbol{\beta}$ is the regression parameter vector, $\boldsymbol{\phi}$ the extra parameters of each distribution (dispersion in most of cases), and $\boldsymbol{\sigma^2} \text{ and } \boldsymbol{\rho}$ are the parameters that compose the variance covariance matrix $\boldsymbol{\Sigma}$. The marginal likelihood function for the model described in \autoref{cap:modelo} for each sample unit is
\begin{equation}
\label{eq:int_likelihood}
\mathrm{L}_{i}(\boldsymbol{\beta},\boldsymbol{\Sigma},\boldsymbol{\phi} \mid \mathrm{\mathbf{y}}) = \int\prod_{r=1}^{k}f(\mathrm{y}_{r}\mid\mathrm{\mathbf{b}}, \boldsymbol{\beta}, \boldsymbol{\phi})f(\mathrm{\mathbf{b}}\mid\boldsymbol{\Sigma})\mathrm{d}\mathrm{\mathbf{b}},
\end{equation}
where $\mathrm{\mathbf{y}}$ is the $k$-response vector, $\mathrm{y_r}$ is each response and $\mathrm{\mathbf{b}}$ is the random effects vector. The full likelihood for $\boldsymbol{\theta}$ is given by

\begin{equation*}
\mathrm{L}(\boldsymbol{\beta}, \boldsymbol{\Sigma}, \boldsymbol{\phi}) = \prod_{i=1}^{N}\mathrm{L}_i(\boldsymbol{\beta},\boldsymbol{\Sigma},\boldsymbol{\phi}\mid\mathrm{\mathbf{y}}),
\end{equation*}
where \emph{N} is the total number of sample units. Under independence between sample units, we have:

\begin{equation}
\label{eq:marginal_likelihood}
\mathrm{\mathbf{L}}(\boldsymbol{\beta}, \boldsymbol{\Sigma}, \boldsymbol{\phi}) = \prod_{i=1}^{N}\int\prod_{r=1}^{k}f(\mathrm{y}_{r}\mid\mathrm{\mathbf{b}}, \boldsymbol{\beta}, \boldsymbol{\phi})f(\mathrm{\mathbf{b}}\mid\boldsymbol{\Sigma})\mathrm{d}\mathrm{\mathbf{b}}.
\end{equation}

The \autoref{eq:marginal_likelihood} is composed of the product of two probability distributions. The first one is the probability distribution of the sample units, while the second is the probability distribution of the random effects. The distribution of the random effects is assumed to be a multivariate normal distribution. As the probability distribution of the response is not normal, the integral does not present a closed form solution expression, thus numerical methods are required to solve such integral.
%
%
\subsubsection{Numerical integration via Laplace approximation}
\label{cap:laplace}

%
%
Once we are dealing with non-normal data, one possibility is to use numerical integration methods to solve the integral in \autoref{eq:int_likelihood} for each set of sample units. Other possibilities include Penalized Quasi Likelihood (PQL) and Monte Carlo techniques that will not be covered here. It is important to note that the integral is of dimension $k$, and $k$ can vary from two to $r$ (the number of response variables). Methods based on numerical quadratures such as trapezium, $1/3$ Simpson, Gauss-Hermite or adaptative Gauss-Hermite use too many points to evaluate the integrand and its complexity is proportional to the dimension of the integral \cite{environmetrics}. To solve the integral in \autoref{eq:int_likelihood} we used the LA, which is a special case of the adaptive Gauss-Hermite quadrature (AGHQ) when it is used only one integration point \cite{laplace}. Although LA is faster than AGHQ, it is less accurate \cite{laplaceAGHQ}.

The idea of LA is to replace an integral with a tractable closed-form expression. This resulting expression is maximized concerning the variable that we wanted to integrate and the integral is solved. Applying the LA to the integral in \autoref{eq:int_likelihood} results in the marginal likelihood function. The approximation is obtained by

\begin{equation}
\label{eq:laplace_approximation}
\int_{\mathbb{R}^k}\text{exp}\{\mathrm{\mathbf{Q(b)}}\}\text{d}\mathrm{\mathbf{b}} \approx (2\pi)^{n_{\mathrm{\mathbf{b}}}/2}\left | -\mathbf{Q''(\hat{b})} \right |^{-1/2}\text{exp}\{\mathbf{Q(\hat{b})}\},
\end{equation}
where $\mathrm{\mathbf{Q(b)}}$ is a uni-modal and bounded function of the variable $\mathrm{\mathbf{b}}$ and $n_\mathrm{\mathbf{b}}$ is the dimension of the integral, that it is $k$ in this case. Careful accounting of the approximation error shows it to generally be $O(n^{-1})$ where $n$ is the sample size (assuming a fixed length for $\mathrm{\mathbf{b}}$) \cite{wood2015}. In this case, $\mathrm{\mathbf{Q(b)}}$ is the log of the integrand in \autoref{eq:int_likelihood}, $\mathrm{\mathbf{b}}$ the random effect, $\mathbf{\hat{b}}$ the maximized random effect values, while all other $\boldsymbol{\theta}$ parameters remain constant. Therefore, $\mathbf{Q(\hat{b})}$ is the maximum value of the function with respect to $\mathrm{\mathbf{b}}$ and $\mathbf{Q''(\hat{b})}$ is the curvature of the function in the maximum, or in other words, it corresponds to the value of the second derivative (Hessian) of the function in the maximum.

It is important to say that the integral of order $k$ in \autoref{eq:int_likelihood} has to be solved repeatedly in every step of the ML method for each one of the $N$ sample units. Therefore, we have two optimizations procedures: one that is called the outer maximization, which is the maximization of the marginal likelihood (result from the integral), and another one that is called the inner maximization, which is the maximization required inside the LA.

For each optimization procedure, we have different situations. In the inner process is feasible to calculate the Hessian matrix efficiently (once we know the integrand) and to choose a good initial guess for $\mathrm{\mathbf{b}}$. Therefore, the inner process maximization can be efficiently implemented by the Newton-Raphson (NR) method (that requires second-order derivatives). On the other hand, in the outer maximization, it is not feasible to calculate the Hessian matrix efficiently, neither we have a good initial guess after the integral is solved. Therefore, it will be maximized with algorithms that require only first-order derivatives and do not depend strongly on the initial guess as NR does. The derivatives were obtained through automatic differentiation \cite{ad}. 



%
%
\section{SOFTWARE IMPLEMENTATION}
\label{cap:software_implementation}
%
%
In this section, we present the software implementation employed to fit the model presented in \autoref{cap:modelo}. The software used was \texttt{R} version 4.0.2 \cite{rsoft} along the Template Model Builder - \texttt{TMB} \cite{tmb} package. \texttt{TMB} is an \texttt{R} package that offers a collection of tools to build complex random effects statistical models through \texttt{C++} templates. \texttt{TMB} is based on state-of-the art software: \texttt{CppAD} \cite{cppad}, \texttt{Eigen C++} \cite{eigen}, \texttt{BLAS} \cite{blas}, among others libraries written in \texttt{C++}, which are responsible for obtaining the derivatives through AD, linear algebra computations and parallelization, respectively. 

Once the user supplies an objective function in a C++ template file (which, usually is the negative log-likelihood function), it obtains the marginal likelihood via LA integrating out the $\mathrm{\mathbf{b}}$ random effects. The internal optimization of the LA is made via Newton's method, with first and second derivatives provided via AD from \texttt{TMB}. After that, the marginal can be optimized to obtain the MLE parameters with any general-purpose algorithm (using first derivative information provided from \texttt{TMB}), such as \texttt{PORT} or \texttt{BFGS} implemented in \texttt{nlminb} and \texttt{optim} routine in R. Moreover, the standard deviation of the parameters (or a function of it) can be obtained via the Delta method \cite{sdreport}; profiling is available too.

A implementation of the Poisson MGLMM is available on http://www.leg.ufpr.br/doku.php/publications:papercompanions.


%
%
\subsection{Reparametrization}
\label{cap:rep}
%
%

We had to use reparametrization techniques for those parameters that vary in a subset of $\mathbb{R}$. Dispersion parameters for both NB and COM-Poisson distributions were \texttt{log} reparametrized (\texttt{ln} to be more precise) into the dispersion parameter as input, allowing it to will vary on all $\mathbb{R}$. For $\boldsymbol{\sigma}$ and $\rho$ we used an efficient reparametrization from \texttt{TMB} under the normal multivariate specification.

Instead of estimating $\boldsymbol{\rho}$ directly in  $\boldsymbol{\Sigma}$, the \texttt{TMB} reparametrization consists to estimate a different unrestricted parameter $\boldsymbol{\varrho}$ in a lower triangular matrix with unit diagonal $L$. An unstructured symmetric positive definite correlation matrix $\boldsymbol{\Omega}$ can be obtained via $\boldsymbol{\Omega} = \mathbf{D}^{-\frac{1}{2}}\mathbf{L}\mathbf{L}^{\top} \mathbf{D}^{-\frac{1}{2}}$, where $\mathbf{D} = \text{diag}(\mathbf{L}\mathbf{L}^{\top})$.

The $\boldsymbol{\Omega}$ matrix is only one part of \texttt{TMB}'s $\boldsymbol{\Sigma}$ decomposition. Consider the case we have 4 response variables, so, we want to estimate 6 correlation parameters. Thus, the lower triangular matrix $\mathbf{L}$ has order 4 and is filled row-wise, such that:
\begin{equation*}
\mathbf{L} = \begin{pmatrix}
1 &  &  & \\
\varrho_0 & 1 &  & \\
\varrho_1 & \varrho_{2} & 1 & \\
\varrho_{3} & \varrho_{4} & \varrho_{5} & 1
\end{pmatrix}.
\end{equation*}

According to the the restrictions imposed into $\boldsymbol{\Omega}$, and the way the normal multivariate was coded into \texttt{TMB}, the variance parameters of the random effects need to be supplied as standard deviations. To obtain the full variance-covariance matrix $\boldsymbol{\Sigma}$ we can use $\boldsymbol{\Sigma} = \mathbf{W}\boldsymbol{\Omega} \mathbf{W}$ where $\mathbf{W}$ is a diagonal matrix with entries being equal to the random effects's standard deviation.


\section{DATA ANALYSIS}
\label{cap:dataanalyses}

In this section we apply the proposed model in \autoref{cap:modelo} to analyze the NHANES data set presented in \autoref{cap:nhanes}. Initial values had to be chosen carefully for the applied models. Firstly, it was fitted a MCGLM model \cite{wbonat.pkg} with \texttt{log} link function, power parameter fixed, and variance Tweedie, which corresponds to fit a Poisson Model via quasi-likelihood, in order to obtain initial parameter estimates for the regression and variance parameters (based on the variance of the residuals). On the other hand, the correlation parameter was set to 0. We did not use MCGLM to obtain initial estimates for the correlation parameter due to the difference in methodologies. This parameter estimates values were considered as initial values to the Poisson model.

Then, for every distribution, the estimation was made in two steps. In the first, the model was fitted based on a random sample (SRS) of size 350. In the second step, the initial values from the sample model were used to fit the model with the whole dataset. After that, the first estimation was carried out with the \texttt{PORT} routine, followed by \texttt{BFGS} and \texttt{PORT}. For the dataset analyzed, \texttt{PORT} was the fastest and produced a more consistent optimization over \texttt{BFGS} in most scenarios. In every step, initial values were used from the last optimization, and the choice of the intermediate models and the final results reported were based on the maximum value of the logLik.

The initial values for the NB model were the final estimates from the Poisson model (with initial value to the dispersion parameter $\phi = 1$, i.e., small overdispersion). In the same way, the initial values for the COM-Poisson model were the final estimates from the NB model (with initial value to the dispersion parameter $\nu = 1$, i.e., equidispersion). We used the mean-parametrization of the COM-Poisson model, where $\nu>1$ is considered underdispersion, and $\nu<1$, overdispersion \cite{compoissonmeanparam}.

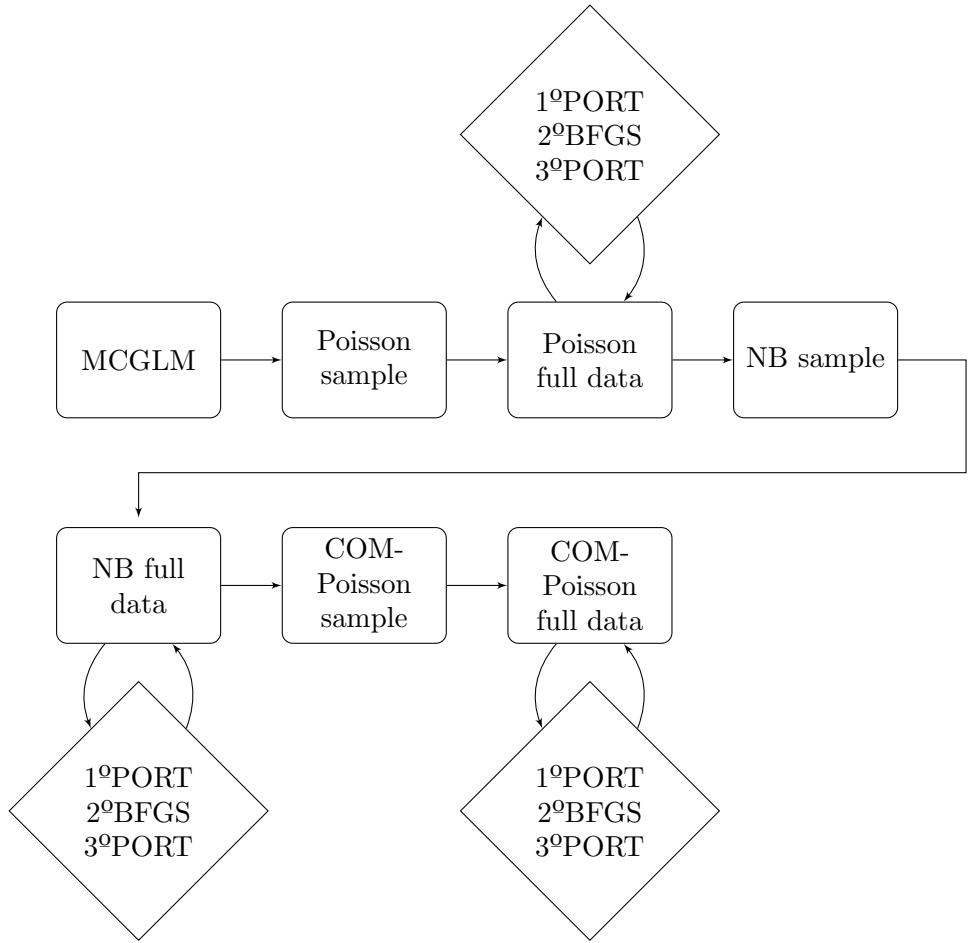
\begin{figure}
\caption{Model estimation process for each dataset}
\centering
\vspace{0.35cm}
\setlength{\abovecaptionskip}{.0001pt}
\begin{tikzpicture}[node distance = 3cm, auto]
    \tikzstyle{decision} = [diamond, draw, fill=white!15, text width=4.5em, text badly centered, 
                             minimum height=2.5em]
    \tikzstyle{block} = [rectangle, draw, fill=white!15, 
        text width=5em, text centered, rounded corners, minimum height=4em]
    \tikzstyle{line} = [draw, -latex']
    \tikzstyle{cloud} = [draw, ellipse,fill=white!20, node distance=3cm,
        minimum height=2em]
    \node [block] (mcglm) {MCGLM};
    \node [block, right of=mcglm] (poissons) {Poisson sample};
    \node [block, right of=poissons] (poissonf) {Poisson full data};
    \node [block, right of=poissonf] (negbins) {NB sample};
    \node [decision, above of=poissonf] (est1) {1ºPORT 2ºBFGS 3ºPORT};
    \node [block, below of=mcglm] (negbinf) {NB full data};
    \node [block, right of=negbinf] (cmps) {COM-Poisson sample};
    \node [decision, below of=negbinf] (est2) {1ºPORT 2ºBFGS 3ºPORT};
    \node [block, right of=cmps] (cmpf) {COM-Poisson full data};
    \node [decision, below of=cmpf] (est3) {1ºPORT 2ºBFGS 3ºPORT};
    \path [line] (mcglm) -- (poissons);
    \path [line] (poissons) -- (poissonf);
    \path [line] (poissonf) -- (negbins);
    \path [line] (negbins) -- node {} +(2,0) |- node {} (0,-1.5) -| node {} (0,-2.1);
    \path [line] (negbinf) -- (cmps);
    \path [line] (cmps) -- (cmpf);
    \path [line] (poissonf) edge[bend left] node [below left] {} (est1);
    \path [line] (est1) edge[bend left] node [below left] {} (poissonf);
    \path [line] (negbinf) edge[bend right] node [below left] {} (est2);
    \path [line] (est2) edge[bend right] node [below left] {} (negbinf);
    \path [line] (cmpf) edge[bend right] node [below left] {} (est3);
    \path [line] (est3) edge[bend right] node [below left] {} (cmpf);
\end{tikzpicture}
\vspace{0.30cm}
\label{fig:estimationprocess}
\end{figure}

This workflow is described in \autoref{fig:estimationprocess}. The reported results are only from the full data, as the samples were used to obtain initial estimates.

\begin{figure}
\caption{Model structure and linear predictor selection}
\centering
\vspace{0.35cm}
\setlength{\abovecaptionskip}{.0001pt}
\begin{tikzpicture}
\usetikzlibrary{shapes,arrows,shadows,decorations.pathreplacing}
\tikzstyle{sensor}=[draw, fill=white!20, text width=8em, 
                    text centered, minimum height=2.5em,drop shadow]
\tikzstyle{wa} = [sensor, text width=7em, fill=white!20, 
                  minimum height=3em, rounded corners, drop shadow]
\tikzstyle{wa2} = [sensor, text width=7em, fill=white!20, 
                   minimum height=3em, rounded corners, drop shadow]
\node (best) [wa2 , align = center ] at (4,0) {Best \\ model};
\begin{scope}[yshift = 3cm]
\node (wa) [wa ]  {Poisson Best};
\draw[-latex] (wa.east) -- (best);
\path (wa.west)+(-3.2,0.5) node (asr1)[sensor ] {Poisson\textsubscript{1}};
\path (wa.west)+(-3.2,-0.5) node (asr2)[sensor ] {Poisson\textsubscript{5}};
\path [draw, ->] (asr1.east) -- node [above ] {} (wa.170);
\path [draw, ->] (asr2.east) -- node [above ] {} (wa.190);
\end{scope}
\begin{scope}[yshift = -0.5cm]
\node (wa) [wa ]  {NB Best};
\draw[-latex] (wa.east) -- (best);
\path (wa.west)+(-3.2,1.5) node (asr1) [sensor ] {NB\textsubscript{1}};
\path (wa.west)+(-3.2,0.5) node (asr2)[sensor ]  {NB\textsubscript{2}};
\path (wa.west)+(-3.2,-0.5) node (asr3)[sensor ] {NB\textsubscript{3}};
\path (wa.west)+(-3.2,-1.5) node (asr4)[sensor ] {NB\textsubscript{4}};
\path (wa.west)+(-3.2,-2.5) node (asr5)[sensor ] {NB\textsubscript{5}};
\path [draw, ->] (asr1.east) -- node [above ] {} (wa.160) ;
\path [draw, ->] (asr2.east) -- node [above ] {} (wa.170);
\path [draw, ->] (asr3.east) -- node [above ] {} (wa.180);
\path [draw, ->] (asr4.east) -- node [above ] {} (wa.190);
\path [draw, ->] (asr5.east) -- node [above ] {} (wa.200);
\end{scope}
\begin{scope}[yshift = -6cm]
\node (wa) [wa ]  {COM-Poisson\\Best};
\draw[-latex] (wa.east) -- (best);
\path (wa.west)+(-3.2,1.5) node (asr1) [sensor ] {COM-Poisson\textsubscript{1}};
\path (wa.west)+(-3.2,0.5) node (asr2)[sensor ]  {COM-Poisson\textsubscript{2}};
\path (wa.west)+(-3.2,-0.5) node (asr3)[sensor ] {COM-Poisson\textsubscript{3}};
\path (wa.west)+(-3.2,-1.5) node (asr4)[sensor ] {COM-Poisson\textsubscript{4}};
\path (wa.west)+(-3.2,-2.5) node (asr5)[sensor ] {COM-Poisson\textsubscript{5}};
\path [draw, ->] (asr1.east) -- node [above ] {} (wa.160) ;
\path [draw, ->] (asr2.east) -- node [above ] {} (wa.170);
\path [draw, ->] (asr3.east) -- node [above ] {} (wa.180);
\path [draw, ->] (asr4.east) -- node [above ] {} (wa.190);
\path [draw, ->] (asr5.east) -- node [above ] {} (wa.200);
\end{scope}
\draw [decorate,decoration={brace,amplitude=1em,mirror,raise=8ex} ]
(asr4.south) --++(4,0) node[midway,yshift=-6.5em, align = center ]{Choosing \\ model \\ structure};
\draw [decorate,decoration={brace,amplitude=1.5em,mirror,raise=2ex} ]
(wa.south) --++(4,0) node[midway,yshift=-4.5em, align = center ]{Linear \\ Predictor \\  Selection};
\node[align=left, anchor=south] (textNode) at (5,2) {1 - Full model \\ 2 - Fixed variance \\ 3 - Comum variance \\ 4 - Fixed dispersion \\ 5 - Correlation zero};
\end{tikzpicture}
\label{fig:linearpredsel}
\end{figure}
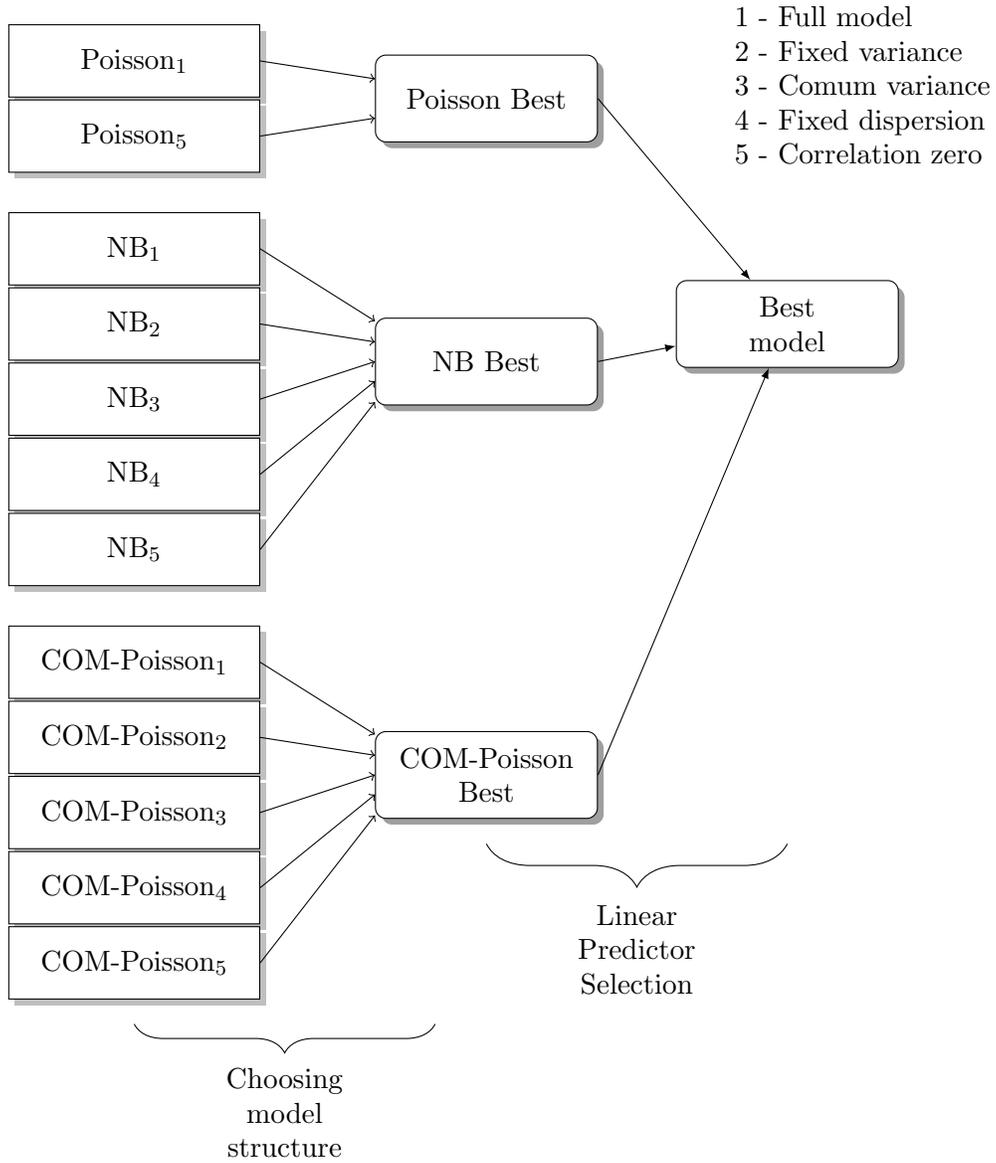

In addition to the models presented in \autoref{cap:modelo}, simpler versions of them were also fitted. We used some different scenarios concerning the variance of the random effects, one for the dispersion parameter and another for the correlation parameter. The first case considered a model with a common variance specification, where the variance of the random effect was equal to all response variables. In the second case, we used a fixed variance specification, where the variance was fixed at $1$. The third scenario was when the dispersion parameter was fixed to $\phi = 1$ for the NB and $\nu=1.5$ for the COM-Poisson model indicating small underdispersion for this distribution. The final and fourth scenario sets the correlation parameter to zero, which aims to reproduce the case where every response distribution would be fitted separately. As the Poisson distribution does not have a dispersion parameter, only the full specification of the model was used and when the correlation parameter was set to zero. This workflow and the linear predictor selection that is explained in the next subsection is presented in \autoref{fig:linearpredsel}.

We reported the results of the models presented in \autoref{fig:estimationprocess} and their variations with log-likelihood value (\texttt{logLik} - larger the best), Akaike Information Criterion (AIC - lower the best), Bayesian Information Criterion (BIC - lower the best), number of parameters estimates (np) and whether the SE was returned or not for the parameter estimates. 


  
\subsection{NHANES data}
\label{cap:resultNHANES}

  
\autoref{tab:nhanesfit} presents the goodness-of-fit measures for NHANES data from different distributions and specifications.

\begin{table}

\caption{Goodenss-of-fit measures for NHANES data from different distributions and specifications}
\centering
\begin{tabular}[t]{lccccc}
\toprule
Model & np & AIC & BIC & Loglik & SE\\
\midrule
Poisson & 21 & 7145.1 & 7253.4 & -3551.6 & \checkmark\\
Rho Zero Poisson & 18 & 7169.8 & 7262.6 & -3566.9 & \xmark\\
\addlinespace
NB & 24 & 7150.2 & 7273.9 & -3551.1 & \checkmark\\
Fixed Dispersion NB & 21 & 8092.4 & 8200.6 & -4025.2 & \xmark\\
Comum Variance NB & 22 & 7203.4 & 7316.8 & -3579.7 & \xmark\\
Fixed Variance NB & 21 & 7924.2 & 8032.5 & -3941.1 & \checkmark\\
Rho Zero NB & 21 & 7176.8 & 7285.1 & -3567.4 & \xmark\\
\addlinespace
COM-Poisson & 24 & 4615.9 & 4739.6 & -2284.0 & \checkmark\\
Fixed Dispersion COM-Poisson & 21 & 6991.7 & 7099.9 & -3474.8 & \xmark\\
Comum Variance COM-Poisson & 22 & 5608.2 & 5721.6 & -2782.1 & \checkmark\\
Fixed Variance COM-Poisson & 21 & 6649.8 & 6758.0 & -3303.9 & \checkmark\\
Rho Zero COM-Poisson & 21 & 5526.3 & 5634.5 & -2742.1 & \checkmark\\
\bottomrule
\end{tabular}
\label{tab:nhanesfit}
\end{table}

Overall, the model which best fitted the data with respect to the three fit measures was the COM-Poisson with full specification. We can also note a large difference in absolute numbers against the competitors. The second best specification was with rho zero for the COM-Poisson model. Moreover, we can note a very close \texttt{logLik} between the NB and the Poisson: it happened due to a large value for the dispersion parameter of the NB, approaching for the Poisson model. Among the smaller specifications, the fixed dispersion was the worst scenario for both NB and COM-Poisson models. 

These results agree with the underdispersion characteristic of the data presented in \autoref{tab:desc_nhanes}. It is important to note that beforehand only COM-Poisson is able to deal with underdispersion data due to the $\nu$ parameter. For Poisson and NB models, the random effects structure only added some extra variance for each response variable. For Poisson, the extra variance parameter gives the flexibility to model only overdispersion; while for NB, it allows modelling even greater variability than the traditional NB model. Therefore, the analysis of this dataset confirmed the expected results by the specified model and distributions.

The checkmark \cmark ~in the last column specifies that the SE of the parameter estimate was returned at least for one parameter of the model. The \xmark ~indicates that it was not possible to return the SE for any parameter estimate of the model. In order to deal with this problem, we modified the linear predictor excluding those covariates that their respective parameter estimates did not have the SEs returned for each response variable based on the best model, and it was refitted. The same linear predictor was used to fit the best model from each distribution, in this case, Poisson, and NB. The results are presented in \autoref{tab:nhanesfit2}.

\begin{table}
\caption{Goodness-of-fit measures for NHANES data from the best specification for each distribution}
\centering
\begin{tabular}[t]{lccccc}
\toprule
Model & np & AIC & BIC & logLik & SE\\
\midrule
Poisson & 16 & 7147.8 & 7230.3 & -3557.9 & \cmark\\
NB & 19 & 7153.1 & 7251.0 & -3557.5 & \cmark\\
COM-Poisson & 19 & 4448.6 & 4546.6 & -2205.3 & \cmark\\
\bottomrule
\end{tabular}
\label{tab:nhanesfit2}
\end{table}

\begin{figure}
\centering
\includegraphics[width=\textwidth]{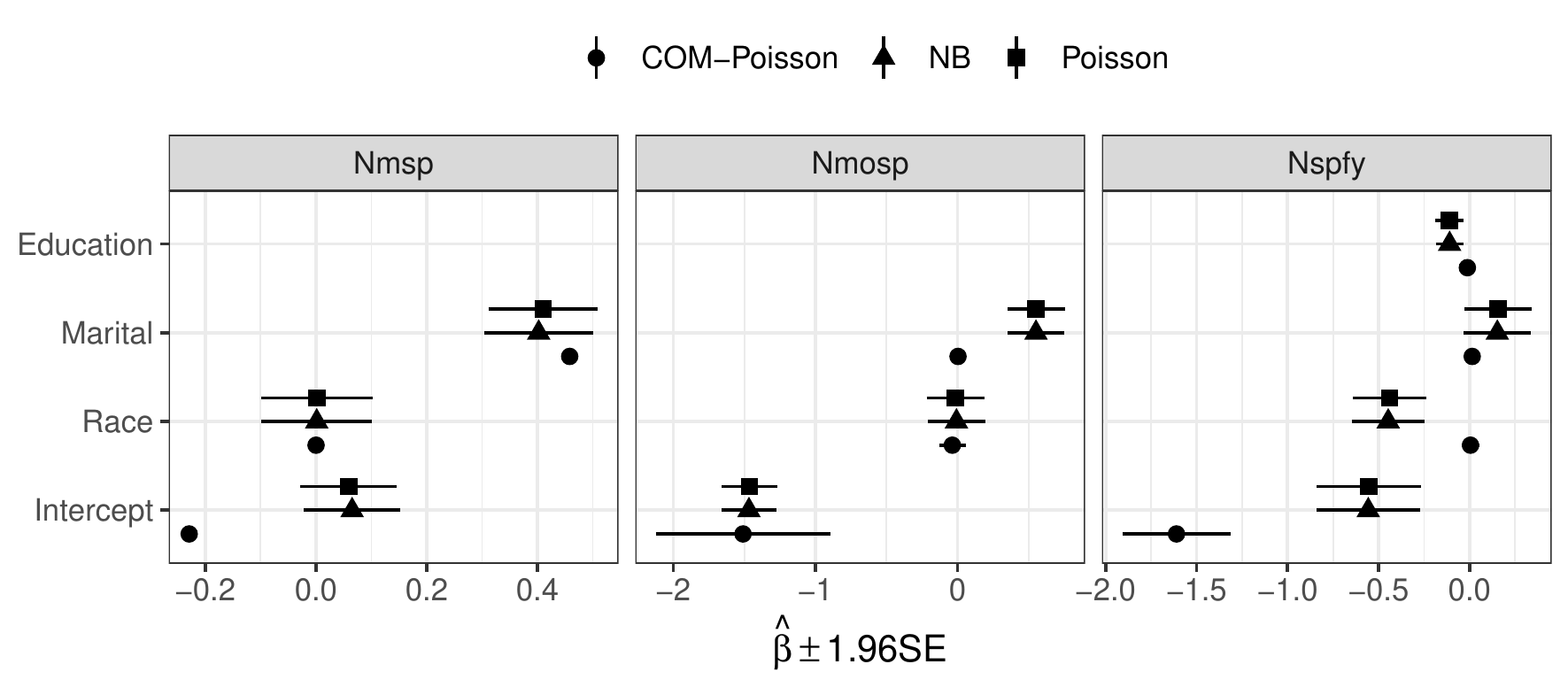}
\caption{Regression parameter estimates and 95\% confidence intervals by outcome and final model}\label{fig:nhanesbeta}
\end{figure}

We can see that reducing the linear predictor improved the COM-Poisson model fit, while it worsened the Poisson and NB in terms of \texttt{logLik} and \texttt{AIC}, but improved the \texttt{BIC} because of a greater penalty due to the number of parameters. In order to better understand the behaviour of the parameter estimates, \autoref{fig:nhanesbeta} compares the regression estimates, \autoref{tab:nhanesdisp} compares the dispersion and \autoref{eq:corrmatrixnhanes} presents a matrix where out of the diagonal are presented correlation estimates and the SE of each random effect, along with the standard deviation and SE in the diagonal.

From \autoref{fig:nhanesbeta} we can see that for most of the estimates, the COM-Poisson confidence intervals were smaller than NB and Poisson models. As these distributions can not handle underdispersed data, they tend to overestimate the SE of the parameter estimates. The confidence interval for Nmosp intercept based on COM-Poisson was the only larger than NB and Poisson models. Moreover, we see that the intercept point estimate was not close between COM-Poisson and the other two models in every scenario.

Regarding the relationship between the covariates and the response variables, the models considered, mostly agree in the direction of the relationship to each response variable. Race (0 = Others, 1 = White) covariate had a null effect for Nmsp and Nmosp; for Nspfy had a negative effect for Poisson and NB models and null for COM-Poisson model. Marital status (0 = Married, 1 = Others) had a positive effect for all three response variables and Poisson and NB models; while COM-Poisson model had only a positive effect for Nmsp. Finally, education level (range from 1 = Less Than $9^{th}$ Grade to 5 = College Graduate or above) had a negative effect on Nspfy for NB and Poisson models and null effect for COM-Poisson. In those models that can not handle underdispersion, more information was captured by the regression parameters than in the COM-Poisson model.

\begin{table}

\caption{Dispersion parameter estimates and SEs for each model and outcome of NHANES data}
\centering
\begin{tabular}[t]{ccccc}
\toprule
\multicolumn{1}{c}{ } & \multicolumn{2}{c}{NB ($\phi$)} & \multicolumn{2}{c}{COM-Poisson ($\hat\nu$)} \\
\cmidrule(l{3pt}r{3pt}){2-3} \cmidrule(l{3pt}r{3pt}){4-5}
Outcome & Estimate & SE & Estimate & SE\\
\midrule
Nmsp & 4958.17 & 21947.1 & 46.513 & 0.059\\
Nmosp & 753.63 & 3878.4 & 20.964 & 5.829\\
Nspfy & 1996.65 & 12993.9 & 19.675 & 1.286\\
\bottomrule
\end{tabular}
\label{tab:nhanesdisp}
\end{table}

From \autoref{tab:nhanesdisp} we can see that $\phi$ was large enough to approximate the NB to a Poisson model followed by an even larger SE; it justifies the similar model measures presented in \autoref{tab:nhanesfit} and \autoref{tab:nhanesfit2}. For COM-Poisson, large $\hat\nu$ values indicate underdispersion and a small SE indicates that $\hat\nu$ is not one, so, it cannot be considered equidispersed.

\begin{equation}
    \begin{aligned}
\boldsymbol{{\Sigma}'_\text{Poisson}} &= 
  \begin{bmatrix}  
    0.18 (0.03)^{*} & 0.94(0.1)^{*} & 0.97 (0.04)^{*}\\ 
    &  0.23(0.08)^{*} & 0.92(0.13)^{*} \\ 
    &   & 0.63(0.07)^{*}
  \end{bmatrix} \\
\boldsymbol{{\Sigma}'_\text{NB}} &=
  \begin{bmatrix}
  0.18 (0.03)^{*} & 0.97(0.08)^{*} & 0.99 (0.02)^{*}\\ 
  &  0.23(0.08)^{*} & 0.97(0.09)^{*}  \\ 
  &   & 0.63(0.07)^{*} 
\end{bmatrix} \\
\boldsymbol{{\Sigma}'_\text{COM-Poisson}}&=
      \begin{bmatrix}  
        0.48 (<0.01)^{*} & <-0.01(<0.01)^{*} & <0.01 (<0.01)\\ 
        &  1.21(0.15)^{*} & <-0.01(0.01) \\ 
        &   & 1.25(0.06)^{*}
      \end{bmatrix} \label{eq:corrmatrixnhanes}
\end{aligned}
\end{equation}

\autoref{eq:corrmatrixnhanes} presents the correlation between random effects in the upper diagonal and the standard deviation in the diagonal. Stars represent statistical significance at 5\% level. It was necessary to make a distinction between $\boldsymbol{\Sigma}$ and $\boldsymbol{{\Sigma}'}$ because the last one uses standard deviation in the diagonal, and the first use variance.

\autoref{eq:corrmatrixnhanes} shows that the correlation estimates for Poisson and NB model are close to one and are significant at 5\% level. In contrast, the correlation estimates for the COM-Poisson model are close to zero and only one is significant (between Nmsp and Nmosp). It seems that the underdispersion information not modelled by Poisson and NB models are somewhat present in the correlation estimates.  Moreover, the standard deviation of the random effect was larger for COM-Poisson than was the Poisson and NB models. It may occur because at the same time that the COM-Poisson has greater variability due to the random effect standard deviation, it balances out with smaller dispersion of $\hat\nu$ parameter.

\section{Discussion}

  The main focus of this article was to propose MGLMM for underdispersed count data. With this new class of statistical models, we can model more than one response variable in the same framework accommodating a random effect that follows a multivariate normal distribution (MN). The parameters of the covariance matrix from the MN allow us to account for the correlation between the random effects and the variance of them. This extra feature is not available in a GLMM model, where it is necessary to model one response variable at a time. 

In particular, we addressed only count data problems, considering Poisson, NB and COM-Poisson distributions. This model was implemented using the \texttt{TMB} package in R. It was estimated under the ML paradigm using numerical integration via LA with inner and outer optimization based on Newton's method and general-purpose algorithm respectively, such as BFGS and PORT routines, which derivatives are provided through AD. 

A dataset from the NHANES survey was analysed by each model and variations of them: rho set to zero, fixed dispersion, fixed variance and common variance for all random effects. It comprised of three response variables, being one equidispersed and two underdispersed. The COM-Poisson was the best model compared to their counterparts according to \texttt{logLik}, \texttt{AIC} and \texttt{BIC}, especially because of its ability to model underdispersion. The SEs were smaller compared to the Poisson and NB models. The dispersion parameter $\nu$ indicated underdispersion, which is expected due to the nature of the data and the correlation parameter $\rho$ was almost zero.

Therefore, we suggest using the MGLMM model framework for count data. In particular, the best results were obtained with the COM-Poisson model in three real datasets. The main advantage of it is the possibility to model all response variables at the same time and measure the correlation between the random effects.

Future work could include the exploration of the residuals for this model and propose a correlation coefficient between the response variables, not the random effects, that it is based on the correlation of the random effect.


%
%
%
%
\section*{Disclosure statement}
No potential competing interest was reported by the authors.
\section*{Funding}
The corresponding author was funded with a scholarship provided by the Coordenação de Aperfeiçoamento de Pessoal de Nível Superior (CAPES).

\bibliographystyle{tfnlm}
\bibliography{Manuscript_with_author_details} 

\end{document}